\documentclass[twocolumn,showpacs,preprintnumbers,amsmath,amssymb, showkeys]{revtex4}

\usepackage{graphicx}
\usepackage{dcolumn}
\usepackage{bm}
\usepackage{epsf}

\begin{document}


\title{Spatial distribution of local density of states in vicinity of impurity on semiconductor surface}

\author{V.\,N.\,Mantsevich}
 \altaffiliation{vmantsev@spmlab.phys.msu.ru}
\author{N.\,S.\,Maslova}%
 \email{spm@spmlab.phys.msu.ru}
\affiliation{%
 Moscow State University, Department of  Physics,
119991 Moscow, Russia
}%

\date{\today }
6 pages, 4 figures
\begin{abstract}
We present the results of detailed theoretical investigations of
changes in local density of total electronic surface states in $2D$
anisotropic atomic semiconductor lattice in vicinity of impurity
atom for a wide range of applied bias voltage. We have found that
taking into account changes in density of continuous spectrum states
leads to the formation of a downfall at the particular value of
applied voltage when we are interested in the density of states
above the impurity atom or even to a series of downfalls for the
fixed value of the distance from the impurity. The behaviour of
local density of states with increasing of the distance from
impurity along the chain differs from behaviour in the direction
perpendicular to the chain.
\end{abstract}

\pacs{71.55.-i}
\keywords{D. Non-equilibrium effects; D. Many-particle interaction; D. Tunneling nanostructures}
\maketitle

\section{Introduction}

    Influence of different impurities on the semiconductor
    local
density of surface states was widely studied experimentally and
theoretically. Most of the experiments were carried out  with the
help of scanning tunneling microscopy/spectroscopy technique \cite
{Dombrowski}, \cite {Boon}, \cite {Kanisawa}. Theoretical
investigations of single impurities and clusters influence on
density of surface states deals with Green's functions formalism
\cite {Sivan}, \cite {Madhavan} or based on the total-energy
density-functional calculations using first-principle
pseudo-potential \cite {Qian}. Numerical calculations based on the
tight-binding model are also carried out \cite {Wiebe}.

    Most of the theoretical calculations don't take into account
modification of local density of continuous spectrum states due to
the influence of impurity atom on semiconductor surface which we
consider to play an important role in the formation of peculiarities
 in the
local density of total surface states .
    So in the present work we suggest a simple model of anisotropic atomic
lattice with impurity atom and we pay special attention to the
changes of local density of continuous spectrum states. This model
suits well for theoretical investigation of $\pi$-bonded chains on
the reconstructed Ge or Si surfaces \cite {Pandey}. It can be also
used for investigation of the sublattices on the cleaved planes of
$A_{III}B_{V}$ semiconductors. We have found that the view of local
density of surface states (amount of downfalls and their shape)
strongly differs depending on the value of the distance from the
impurity atom position and from the direction of observation (along
the atomic chain or perpendicular to the atomic chain).
    It will be shown that downfall in
the resonance when we are interested in the density of states above
the impurity atom can transforms to a series of downfalls or even to
a peak for different values of the distance from the impurity.
\section{The suggested model and main results}
 We shall analyze $2D$ anisotropic atomic lattice formed by the similar atoms with
energy levels $\varepsilon_{1}$ and similar tunneling transfer
amplitudes between the atoms $t$ along the atomic chain. The
interaction between atomic chains is described by tunneling
amplitude $T$, which has the same value for all the similar atoms in
the chain (Fig.~1). Distance between the atoms in the atomic chain
is equal to $a$, distance between the atoms in the neighboring
chains is equal to $b$. Atomic lattice includes impurity atom with
energy level $\varepsilon_{d}$, tunneling transfer amplitude from
impurity atom to the nearest atoms in the atomic chain $\tau$ and to
the nearest atoms in the neighbor chains $\Im$ .

    The model system can be described by the Hamiltonian: $\Hat{H}$:

$$\Hat{H}=\Hat{H}_{0}+\Hat{H}_{imp}+\Hat{H}_{tun}$$
\begin{eqnarray}
&\Hat{H}_{0}&=\sum_{i}\varepsilon_{i}c_{i}^{+}c_{i}+\sum_{<i,j>}t c_{i}^{+}c_{j}+\sum_{<k,l>}T c_{k}^{+}c_{l}+h.c.\nonumber\\
&\Hat{H}_{tun}&=\sum_{i,d}\tau c_{i}^{+}c_{d}+\sum_{k,d}\Im c_{k}^{+}c_{d}+h.c.\nonumber\\
&\Hat{H}_{imp}&=\sum_{d}\varepsilon_{d}c_{d}^{+}c_{d}\nonumber\\
\end{eqnarray}
    $\Hat{H}_{0}$ is a typical Hamiltonian for atomic lattice with hoppings without any impurities.
 $\Hat{H}_{tun}$ describes transitions between
impurity atom and neighboring atoms of the atomic lattice.
$\Hat{H}_{imp}$ corresponds to the electrons in the localized state
formed by the impurity atom in the atomic chain.

Indexes i,j correspond to the direction along the chain; indexes k,l
correspond to the direction perpendicular to the chain.

\begin{figure}[h]
\centering
\includegraphics[width=70mm]{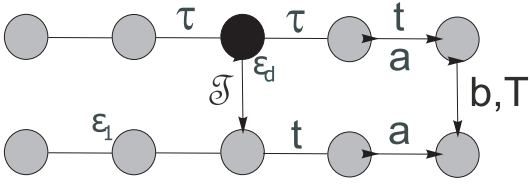}%
\caption{ Schematic diagram of $2D$ atomic lattice with impurity
atom.}
\end{figure}

    We shall use diagramm technique in our
investigation of atomic chain local density of states.

    The dependence of local density of states on the distance along the
atomic chain and in the direction perpendicular to the atomic chain
in the presence of impurity atom is described by the equation:
\begin{eqnarray}
 \rho(\omega,\vec r)=
\frac{-1}{\pi}Sp \Big( \; {\rm Im}\sum_{\vec{\kappa},\vec{\kappa_1}}
\hat{G}^{R}(\vec{\kappa},\vec{\kappa_1},\omega)e^{i\vec{\kappa}\vec
r}
 e^{i\vec{\kappa_1}\vec r} \; \Big)
 \label{LDOS_equ}
\end{eqnarray}
    Where $\vec{r}=(x,y)$, $\vec{\kappa}=(\kappa_{x},\kappa_{y})$ and
$\vec{\kappa_{1}}=(\kappa_{x1},\kappa_{y1})$.
    Green function $\hat{G}^{R}(\vec{\kappa},\vec{\kappa_1},\omega)$
corresponds to the electron transition from the impurity to the
semiconductor continuum states and can be found from the system of
equations:

\begin{multline}
{G}^{R}_{{\kappa_x}0dd}={G}^{0R}_{{\kappa_x}0{\kappa_x}0}\tau{G}^{R}_{dd}
+{G}^{0R}_{{\kappa_x}0{\kappa_x}0} T\sum_{k_y}{G}^{R}_{{\kappa_x}{\kappa_y}dd}\nonumber\\
{G}^{R}_{0{\kappa_y}dd}={G}^{0R}_{0{\kappa_y}0{\kappa_y}}\Im{G}^{R}_{dd}
+{G}^{0R}_{0{\kappa_y}0{\kappa_y}} t\sum_{k_x}{G}^{R}_{{\kappa_x}{\kappa_y}dd}\nonumber\\
{G}^{R}_{dd}={G}^{0R}_{dd}+{G}^{0R}_{dd}\tau\sum_{k_x}{G}^{R}_{{\kappa_x}0dd}
+{G}^{0R}_{dd}\Im\sum_{k_y}{G}^{R}_{0{\kappa_y}dd}\nonumber\\
{G}^{R}_{\vec{\kappa}dd}={G}^{0R}_{\vec{\kappa}\vec{\kappa}}\tau{G}^{R}_{{\kappa_x}0dd}+{G}^{0R}_{\vec{\kappa}\vec{\kappa}}\Im{G}^{R}_{0{\kappa_y}dd}\nonumber\\
{G}^{R}_{{\vec\kappa\vec\kappa_1}}={G}^{0R}_{{\vec\kappa\vec\kappa}}+{G}^{0R}_{{\vec\kappa\vec\kappa}}\tau{G}^{R}_{dd\vec{\kappa_1}}+{G}^{0R}_{{\vec\kappa\vec\kappa}}\Im{G}^{R}_{dd\vec{\kappa_1}}\nonumber\\
\end{multline}

Where zero Green function is evaluated for the $2D$ atomic lattice
without any impurities and has the form:

\begin{eqnarray}
{G}^{0R}_{{\kappa_x}{\kappa_y}}(\omega)=\frac{1}{\omega-\varepsilon_{1}-2t\cdot\cos(k_{x} a)-2T\cdot\cos(k_{y}b)}\nonumber\\
\end{eqnarray}

    Substituting the expression for  Green function
$\hat{G}^{R}(\vec{\kappa},\vec{\kappa_1},\omega)$ obtained from
system  into equation (~\ref{LDOS_equ}) and performing summarization
over wave vectors $k_{x}$ and $k_{x1}$ ($k_{y}$ and $k_{y1}$) we get
the final expression for the local density of continuous spectrum
states along (perpendicular) the atomic chain $\rho_{volume}(x)$
($\rho_{volume}(y)$):
\begin{eqnarray}
\rho_{volume}(\omega,x)=\rho_{0}(\omega)\cdot\frac{(\omega-\varepsilon_{d})^{2}+\gamma^{2}\cdot(1-f(2k_{x}(\omega)x)}{(\omega-\varepsilon_{d})^{2}+\gamma^{2}}\nonumber\\
\rho_{volume}(\omega,y)=\rho_{0}(\omega)\cdot\frac{(\omega-\varepsilon_{d})^{2}+\gamma^{2}\cdot(1-f(2k_{y}(\omega)y)}{(\omega-\varepsilon_{d})^{2}+\gamma^{2}}
\end{eqnarray}

    Where parameter $\gamma=(\tau^{2}+\Im^{2})\cdot\rho_{0}(\omega)$ corresponds to relaxation rate of
electron distribution at the localized state formed by impurity
atom, $\rho_{0}(\omega)$ is a local density of states for the atomic
chain without any impurities. Functions $f(2k_{x}(\omega)x)$ and
$f(2k_{y}(\omega)y)$ are periodical and have the property:
$f(2k_{x}(\omega)x)=1$ if $x=0$ and $f(2k_{y}(\omega)y)=1$ if $y=0$.
If both directions are equivalent
$f(2k_{x}(\omega))=f(2k_{y}(\omega)y)=
J_{0}(2k_{x}(\omega))$.Expression for $k_{x}(\omega)$ or
$k_{y}(\omega)$ can be found from the dispersion law of the $2D$
atomic lattice which has the form.
\begin{eqnarray}
\omega(k_{x},k_{y})=2t\cdot\cos(k_{x}a)+2T\cdot\cos(k_{y}b)\nonumber\\
\end{eqnarray}

    Impurity atom density of states has lorentzian form line shape and can be
evaluated as:

\begin{eqnarray}
\rho_{impurity}(\omega)=-\frac{1}{\pi}\cdot
Im\sum_{d}G_{dd}^{R}(\omega)=\frac{\gamma^{2}}{(\omega-\varepsilon_{d})^{2}+\gamma^{2}}\nonumber\\
\end{eqnarray}

 Local density of total surface states is the result of summarization
 between
local density of continuous spectrum states and impurity atom
density of states.

\begin{eqnarray}
\rho(\omega)=\rho_{volume}(\omega)+\rho_{impurity}(\omega)\nonumber\\
\end{eqnarray}
Let's start from the $1D$ case of the atomic chain. In this case it
is necessary to put in the Hamiltonian: $b=0$, $T=0$ and $\Im=0$.
Final expression for the local density of continuous spectrum states
$\rho_{volume}$ will have the form:
\begin{eqnarray}
\rho_{volume}(\omega)=\rho_{0}(\omega)\cdot\frac{(\omega-\varepsilon_{d})^{2}+\gamma^{2}\cdot(1-\cos(2k(\omega)r)}{(\omega-\varepsilon_{d})^{2}+\gamma^{2}}\nonumber\\
\label{LDOS_equ_1}
\end{eqnarray}
Expression for $k(\omega)$ can be found from the dispersion law of
the $1D$ atomic chain.

\begin{figure*}
\centering
\includegraphics[width=180mm]{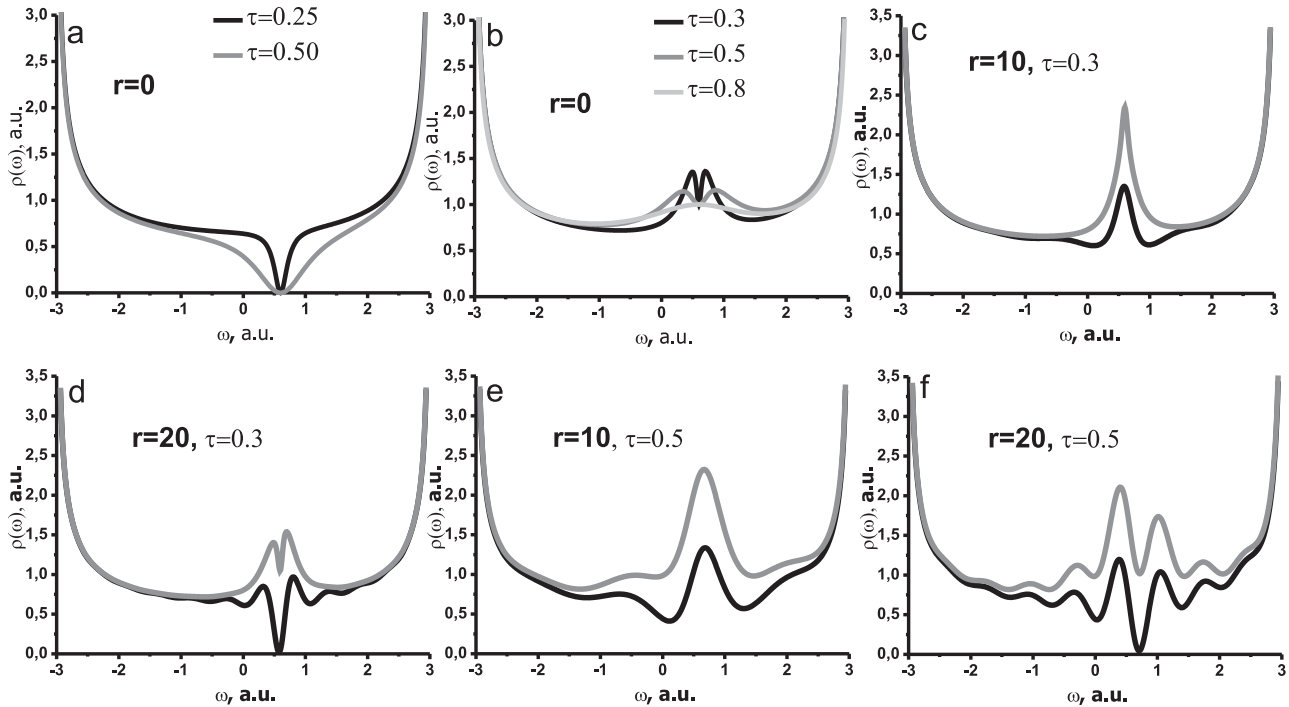}%
\caption{ a) Local density of continuous spectrum states in the case
of the distance from the impurity atom along the atomic chain equal
to zero. b) Local density of surface spectrum states in the case of
the distance from the impurity atom along the atomic chain equal to
zero. c)-f) Local density of continuous spectrum states (black line)
and local density of surface spectrum states (grey line) for the
different values of the distance from the impurity atom along the
atomic chain. For all the figures values of the parameters $a=1$,
$t=1,5$, $\varepsilon_{d}=0,6$ are the same.}
\end{figure*}
  Typical
numerical results for local density of total surface states and
local density of continuous spectrum states calculated above the
impurity atom in $1D$ case (distance value is equal to zero $(r=0)$)
are shown on (Fig.~2a,b). For the local density of continuous
spectrum states (Fig.~2a) a downfall exist in the resonance when
energy is equal to the impurity atom energy level deposition
$(\omega=\varepsilon_{d})$. Width of the downfall depends on the
parameters of the atomic chain, such as relaxation rate or tunneling
transfer amplitude, it rises with the increasing of tunneling
transfer amplitude from impurity atom to the neighbor atoms of the
atomic chain.
     Local density of
surface states (Fig.~2b) has lorentzian form with a downfall in the
resonance. With the increasing of relaxation rate (increasing of
$\tau$) the downfall depth at the top of the peak decreases,
resonance peak shape spreads and it's amplitude falls down.

    Now let's start to analyze the dependence of local density of
continuous spectrum states and local density of surface states at
the fixed value of the distance $r$ along the atomic chain from the
impurity atom position (Fig.~2c-f). We shall again start from the
local density of continuous spectrum states (black lines on Fig.~2).

    When the value of a distance is not equal to zero a series of downfalls in
the local density of continuous spectrum exists. Amount of downfalls
increases with the increasing of distance value and downfalls
amplitude decreases when energy aspire to the edges of the band.
 The most significant amplitude of the downfalls
corresponds to the vicinity of the resonance region. It is clearly
evident that positions of the downfalls on the energy scale can be
found from the equation $2\pi n=2k(\omega)r$ where $n$ is an integer
number. This means that numerator of the equation
(~\ref{LDOS_equ_1}) is equal to zero. When the distance is not equal
to zero not only a downfall in the resonance (Fig.~2d,f) but also a
peak (Fig.~2c,e) can exist in the local density of continuous
spectrum states. At the fixed parameters of the atomic chain
existance of a downfall or a peak in the resonance is determined by
the value of the distance.
\begin{figure*}
\centering
\includegraphics[width=180mm]{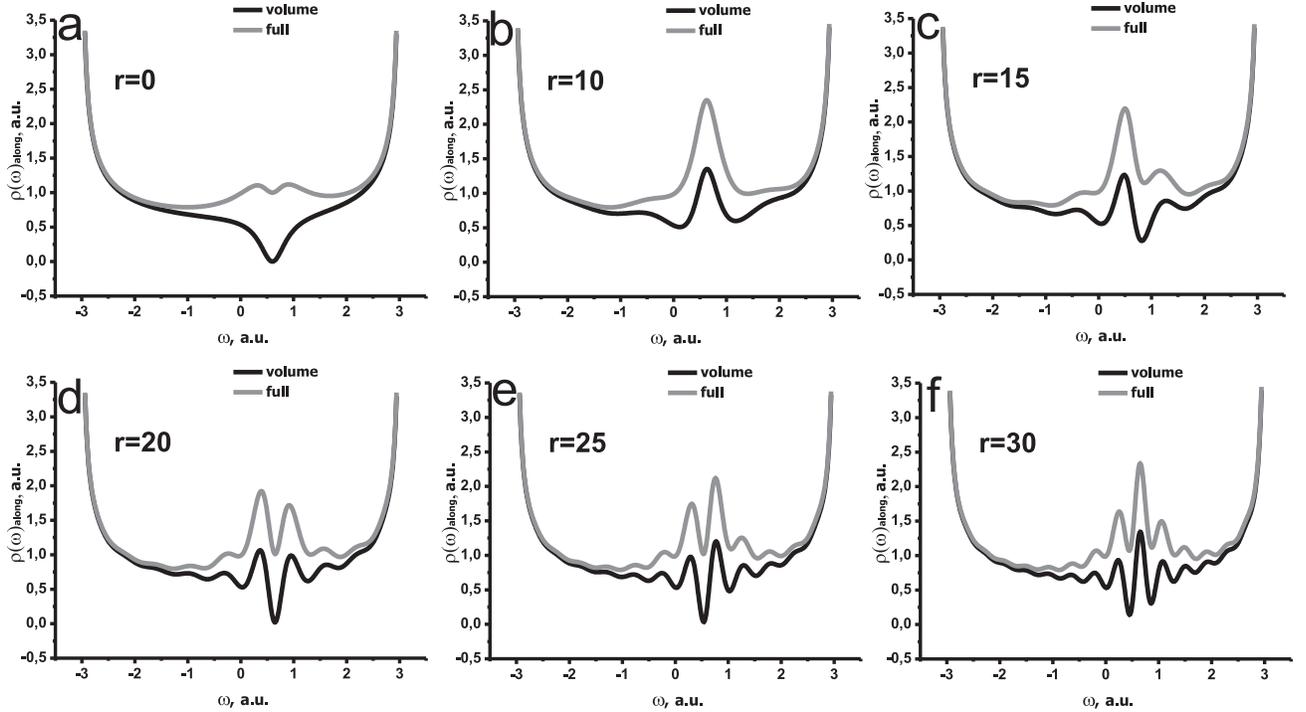}%
\caption{ a) Local density of continuous spectrum states (black
line) and local density of surface spectrum states (grey line) for
the different values of the distance from the impurity atom along
the atomic chain. For all the figures values of the parameters
$a=1$, $b=2$, $t=1,5$, $T=1,2$, $\tau=0,6$, $\Im=0,3$,
$\varepsilon_{d}=0,6$ are the same.}
\end{figure*}
    Local density of surface states is shown by the grey line on Fig.~2c-f.
Comparison between local density of surface states and local density
of continuous spectrum states for the fixed value of $r$ make it
clearly evident that impurity atom density of states can drastically
influence on the local density of continuous spectrum states. Result
depends on the value of tunneling transfer amplitude from impurity
atom to the nearest atoms of the chain. We have found that the most
significant influence corresponds to the situation when tunneling
transfer amplitude between the atoms in the chain $t$ significantly
exceeds tunneling transfer amplitude from impurity atom to the atoms
of the chain $\tau$.  In this case downfall in the resonance becomes
a peak with a small downfall on the top of the peak (Fig.~2e). So
only one significant downfall exists in the resonance region and
there are no downfalls at the energies different from the resonance
value (Fig.~2c)in the local density of surface states, while in the
continuous spectrum density of states a series of downfalls exists.
With increasing of $\tau$ local density of surface states differs
from the local density of continuous spectrum states only by the
amplitude of the downfalls (Fig.~2e,f). In this case number of
downfalls is the same in comparison with the continuous spectrum
density of states, downfalls don't change their shape or position on
the energy scale and contribution to the local density of surface
states from the impurity atom can be considered to be a background.
This effect can be qualitatively understood by the following way:
with increasing of transfer amplitude from impurity atom to the atom
of the chain increases relaxation rate and lorentzian form peak of
impurity atom density of states spreads and it's amplitude
decreases.

 Now let's start to analyze $2D$
atomic lattice. Numerical results for local density of total surface
states and local density of continuous spectrum states calculated
above the impurity atom in perpendicular directions (along the
atomic chain and perpendicular to the atomic chain) are shown on
Fig.~3a, Fig.~4a. It's clear that in this case all the results are
equal for both directions and downfall in the continuous spectrum
density of states or a peak in the local density of surface states
poses just the same properties as in the case of $1D$ atomic chain.

   Let's analyze the situation when the value of the
distances from the impurity atom position  along the atomic chain
 (Fig.~3b-f) and perpendicular to the atomic chain (Fig.~4b-f)
are not equal to zero. We shall start from the local density of
continuous spectrum states (black lines on Fig.~3,4).

  In this case number of downfalls,their shape and position on the energy
scale in each of the perpendicular directions can be found from the
dispersion law just in the same way as for $1D$ atomic chain.
 When the distance
is not equal to zero not only a downfall in the resonance
(Fig.~3d,e; Fig.~4e,f) but also a peak (Fig.~3c,f; Fig.~4c,d) can
exists in the local density of continuous spectrum states.

    \begin{figure*}
\centering
\includegraphics[width=180mm]{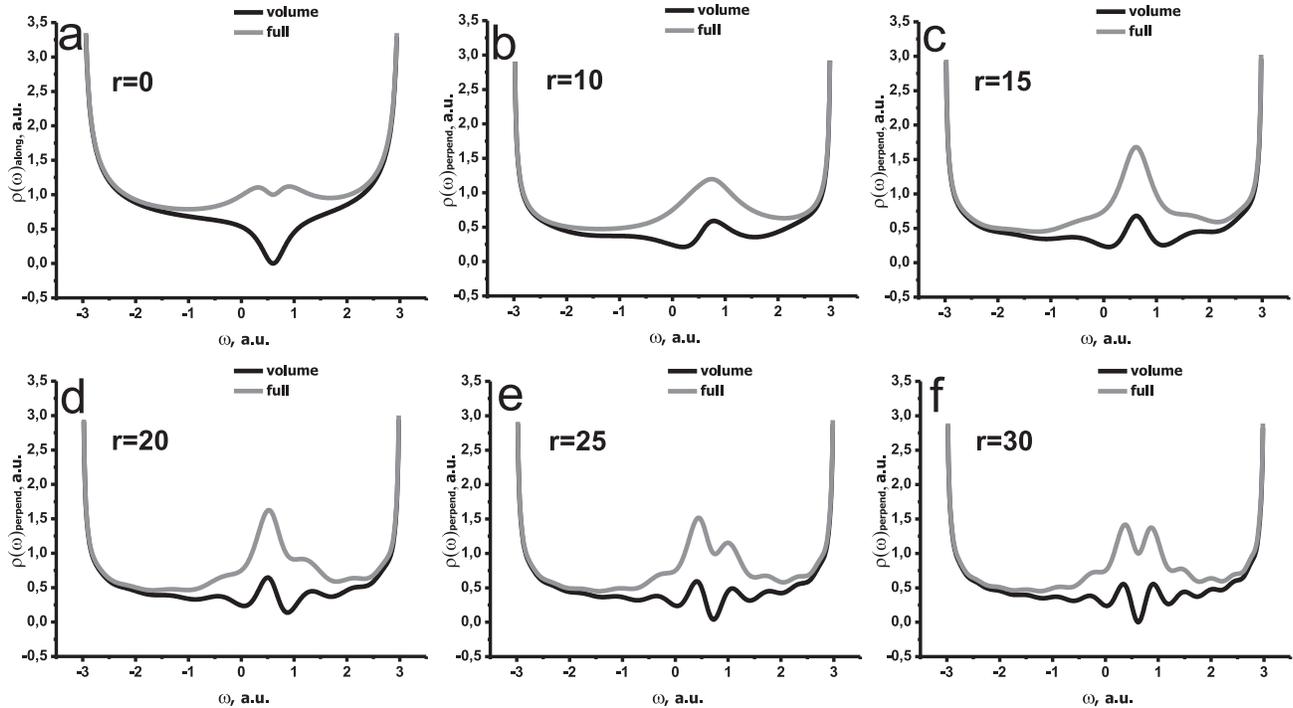}%
\caption{ Local density of continuous spectrum states (black line)
and local density of surface spectrum states (grey line) for the
different values of the distance from the impurity atom
perpendicular to the atomic chain. For all the figures values of the
parameters $a=1$, $b=2$, $t=1,5$, $T=1,2$, $\tau=0,6$, $\Im=0,3$,
$\varepsilon_{d}=0,6$ are the same.}
\end{figure*}
 We have found distance interval for both directions
(along the atomic chain and perpendicular to the atomic chain) where
exists replacement of a peak by a downfall (Fig.~3c-f; Fig.~4c-f) in
local density of continuous spectrum states (and also in the local
density of surface states). Moreover peak in one direction can
corresponds to a downfall in the another direction. Let's analyze
this interval carefully. We shall start from the distance value when
peaks in the resonance in local density of continuous spectrum
states for both directions can be seen ($x(y)=15\cdot a$) (Fig.~3c;
Fig.~4c). With increasing of the distance value ($x(y)=20\cdot a$) a
resonance peak in the direction perpendicular to the atomic chain
still exists (Fig.~4d) and in the direction along the atomic chain a
downfall appears (Fig.~3d). Further increasing of the distance value
($x(y)=25\cdot a$) shows that in the direction along the atomic
chain a downfall still exists (Fig.~3e) and in the perpendicular
direction a downfall substitutes peak (Fig.~4e). Finally when the
value of distances in both directions becomes equal to $x(y)=30\cdot
a$ in the perpendicular direction a downfall still exists (Fig.~4f)
and in the direction along the atomic chain a resonance peak can be
seen (Fig.~3f).

    Local density of total
surface states is shown by the grey lines on Fig.~3,4. It is clearly
evident that for the studied parameters of the system taking into
account impurity atom density of states slightly changes local
density of total surface states in comparison with local density of
continuous spectrum states for both directions. In this case number
of downfalls is the same in comparison with the continuous spectrum
density of states, downfalls don't change their shape or position on
the energy scale.

\section{Conclusion}
 In this work we have shown
that taking into account changes of the local density of continuous
spectrum states formed by the presence of the impurity atom in the
$2D$ anisotropic atomic lattice or even in the $1D$ atomic chain
leads to significant modification of the total local density of
surface states and consequently to the modification of STS spectra.
    We have found that a downfall exists in the STS spectra measured
just above the impurity when impurity atom energy level is equal to
the applied bias voltage. With changing of the distance from the
impurity a series of downfalls is formed on the energy scale both in
the local density of continuous spectrum states and in the local
density of total surface states. Number of downfalls and their
position are determined by the atomic lattice dispersion law. It was
shown that at some values of the distance from the impurity a peak
can exist in the resonance region instead of a downfall. We have
found that behaivour of  local density of surface states depends on
the direction of the observation. Switching on and off of impurity
atom in both directions was found. This effect can be well observed
experimentally with the help of STM/STS technique.

This work was  supported by RFBR grants and by the National Grants
for technical regulation and metrology $01.648.12.3017$ and
$154-6/259/4-08$.


\pagebreak

\end{document}